# GPTutor: Great Personalized Tutor with Large Language Models for Personalized Learning Content Generation

Eason Chen
eason.tw.chen@gmail.com
Carnegie Mellon University

Jia-En Lee
Carnegie Mellon University

Jionghao Lin
Carnegie Mellon University

Kenneth Koedinger
Carnegie Mellon University

## ABSTRACT
We developed GPTutor, a pioneering web application designed to revolutionize personalized learning by leveraging the capabilities of Generative AI at scale. GPTutor adapts educational content and practice exercises to align with individual students' interests and career goals, enhancing their engagement and understanding of critical academic concepts. The system uses a serverless architecture to deliver personalized and scalable learning experiences. By integrating advanced Chain-of-Thoughts prompting methods, GPTutor provides a personalized educational journey that not only addresses the unique interests of each student but also prepares them for future professional success. This demo paper presents the design, functionality, and potential of GPTutor to foster a more engaging and effective educational environment.

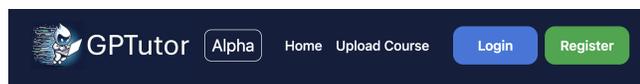

**Figure 1: Screenshot of our application (https://gptutor.academy). In the screenshot, GPTutor personalized the linear regression teaching content with the analogy of the character screen time and the popularity of the famous anime Jujutsu Kaisen.**

## 1 INTRODUCTION TO GPTUTOR

In this demo paper, we introduce GPTutor, a web application designed to generate personalized, analogy-driven educational content derived from source material supplied by educators and tailored to the student's interests and career goals. GPTutor is deployed using a serverless architecture, allowing it to serve many students at scale.

Personalized learning that aligns educational content with students' career goals could not only enhance engagement and efficiency [1, 5, 6] but also empower students to apply the knowledge they learn in their future careers [7, 8]. Nevertheless, creating such personalized learning content is challenging and time-consuming.

In light of recent advances in generative AI (GenAI), particularly the Large Language Models for content generation, AI educational products continue to emerge. Nevertheless, there is not yet an educational content management system that allows teachers to upload their course materials so that students can engage in personalized learning using LLMs. We created GPTutor to fill this gap.

Through GPTutor, neither students nor teachers need to worry about setting up the prompt or ensuring they input the correct learning content in the prompt. This advancement significantly boosts the accessibility of AI-driven personalized learning on a large scale. Teachers can easily upload their curriculum, course contents, and exercises through GPTutor Course Editor. Then, when students learn from these courses, they can input their interests and career goals, and the LLM will customize the course content based on these inputs.

## 2 SYSTEM DESIGN AND DEVELOPMENT

After entering GPTutor, users can select a course and then input their personal interests to personalize the course. GPTutor integrates with OpenAI's GPT-4 API to customize the course curriculum to align with the student's interests.

Students can then start their learning journey. Once they access the course content, the GPT-4 API generates personalized learning content and practice based on the student's interests, the original course material, and the tailored curriculum.

Inspired by the Chain of Thought approach [9], GPTutor generates the curriculum first and then the course content to enhance quality. This is necessary because the course is divided into several sections and subsections that progressively introduce a topic. For example, when teaching linear regression, a teacher usually first explains independent and dependent variables and then describes the concept of slope. The scope and learning goals need to be clearly defined in the curriculum to prevent hallucinations. Moreover, as examples used in the course often span multiple subsections, generating a coherent curriculum ensures that the subsequent learning content is cohesive.

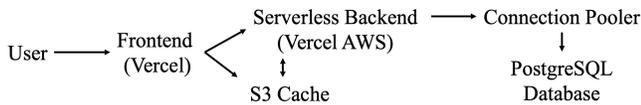

Figure 2: Topic-related practices generated by GPTutor based on the Chain of Thought method.

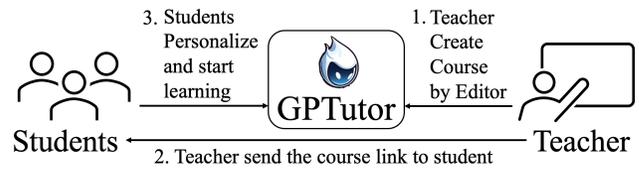

Figure 3: GPTutor's architecture, which utilizes AWS's Serverless framework combined with S3 for caching and PostgreSQL as the database.

As shown in Figure 3, GPTutor is hosted using NextJS and deployed via Vercel AWS's serverless architecture. When users request personalizedization, the serverless backend responds instantly, sending requests to OpenAI's backend using an API Key and the prompt, and then delivering the generated results. This setup ensures the privacy of the Prompt and API Key, preventing misuse. The serverless backend can scale instantly to accommodate a theoretically unlimited number of students concurrently using the system under OpenAI's rate limits. In our stress testing, the system can handle up to 100 simultaneous calls without issues.

## 3 INTERACTION FLOW

Please see the demo video for GPTutor at https://youtu.be/kK078yABHmQ and try it yourself at https://gptutor.academy. The interaction flow is depicted at Figure 4. A teacher who wants to create a course at GPTutor will perform the following:

(1) **Log In:** Teachers begin by logging into GPTutor and clicking on the "Upload Course" button at the upper right corner.

Figure 4: Illustration of how teacher and students will interact with GPTutor.

(2) **Start Editing:** At this stage, teachers have the option to import course content using JSON or CSV formats. They can also start editing a course from scratch.
(3) **Edit Course by GPTutor Editor:** Once the content is uploaded or created, teachers move to the interactive GPTutor Editor for further editing. In the editor, teachers can adjust the summary, scope, and learning goals for each section.
(4) **Provide More Context:** Teachers can also optionally add example course content and practice exercises. These example contents will be used to guide the AI to generate personalized learning content for students.

Then, a student will perform the following:

(1) **Visit the course website:** The student opens the course link provided by the teacher.
(2) **Personalization:** The student selects "Personalize" and enters their interests or career goals.
(3) **Generate Curriculum:** The AI begins to generate a personalized curriculum, which will take about two minutes.
(4) **Curriculum Review and Save:** Once the AI has generated the curriculum, the student clicks "Save and Start Learning."
(5) **Content Creation:** The AI creates personalized learning content based on the student's personalized curriculum and the example teaching content provided by the teacher.
(6) **Start Learning:** Student start learning the content. As illustrated in Figure 1.

## 4 FUTURE WORKS

In the future, we hope to integrate GPTutor with existing work to empower users to interact with the personalized content [3] and further address their questions or audit the contents by GPTutor. Additionally, we aim to explore the best design for generating results to help users grasp the learning goals effectively [2, 4].

## ACKNOWLEDGMENTS

This work is supported by funding from the U.S. National Science Foundation with grants 1740775 and 2213791.